\documentclass[12pt]{article}
\usepackage{graphicx}

\begin{document}
%
\title{Feedback loops of attention in peer production}

\author{Fang Wu, Dennis M. Wilkinson, and Bernardo A. Huberman\\
HP Labs, Palo Alto, California 94304}

\maketitle

\begin{abstract}
A significant percentage of online content is now published and consumed via the mechanism of crowdsourcing. While any user can contribute to these forums, a disproportionately large percentage of the content is submitted by very active and devoted users, whose continuing participation is key to the sites' success.  As we show, people's propensity to keep participating increases the more they contribute, suggesting motivating factors which increase over time. This paper demonstrates that submitters who stop receiving attention tend to stop contributing, while prolific contributors attract an ever increasing number of followers and their attention in a feedback loop. We demonstrate that this mechanism leads to the observed power law in the number of contributions per user and support our assertions by an analysis of hundreds of millions of contributions to top content sharing websites \texttt{Digg.com} and \texttt{Youtube.com}.
\end{abstract}

\section{Introduction}
Websites allowing Internet users to produce, share, rate and classify content have become increasingly popular in the last decade and now represent many of the Internet's most frequently visited pages\footnote{\texttt{http://www.alexa.com/topsites}.}. The most successful of these efforts have experienced rapid growth and attained mainstream popularity, while many others have stayed small or disappeared.

It has been observed that the distribution of the number of contributions per user in many popular peer production efforts follow a heavy-tailed power law form, in which the most active users contribute a disproportionately large percentage of the content \cite{wilkinson-08}. Indeed, these active members often demonstrate extraordinary devotion, contributing for many months or more at an elevated rate, and are often responsible for much of the popular content. Their continuing participation is therefore critical to the success of a peer production effort.

The power law distribution in contributions per user indicates that active users are more likely to continue the more they contribute \cite{wilkinson-08}, suggesting incentives which increase over time. In previous survey-based studies, users cited motivations such as enjoyment, intellectual stimulation, idealism, and a sense of belonging as important to why they participated \cite{shah,lakhani,nov,hsu,oreg}. While these factors no doubt carry weight in a complex way for many users \cite{roberts}, it seems unlikely that they continue to strengthen after weeks or months of contributing. And while status and recognition play a key role in some intimate online communities \cite{hsu,oreg,lampel,okoli}, they are less relevant in one-to-many content-sharing forums which nevertheless display the same power law in the distribution of contributions.

Getting attention paid to one's contributions is a form of value \cite{franck}; people are willing to forsake financial gain for it \cite{huberman-04}. Attention was previously shown to spur further contributions in video sharing \cite{romero} and blogging \cite{miura}. However, no attempt was made to account for social effects in how attention was allocated among the various contributors in this work.

This paper addresses how attention and followers affect contributions to peer production websites, and in particular how they motivate prolific contributors to remain active for a long time. We first demonstrate a strong correlation between lack of attention and users' decision to stop contributing in both \texttt{Youtube.com} and social news site \texttt{Digg.com}. Next, we show that the more users contribute, the greater their following (as measured by their number of digg ``fans'' or youtube ``subscribers''), and the more attention they tend to receive from these followers. Finally, we present a mechanism incorporating these empirical findings which explains the observed power law in the distributions of contributions per user. Taken together, our findings suggest that the feedback loop of attention is an important ingredient in a successful of a peer production effort because of its role in motivating devoted contributors to persist.

\section{\label{sec:data}Data}

In order to test our theory we collected data from \texttt{Digg.com} and \texttt{Youtube.com}, the foremost social news and video sharing websites respectively. In both cases, users submit content that can be consumed by other users, either in the form of a link to news stories (\texttt{Digg}) or a video (\texttt{Youtube}).\footnote{In general, all users can see all items; \texttt{Youtube} does allow people to make their videos private or password protected, but this service is rarely used in practice.} In both systems, attention is transparent and easily observable, and the users are allowed to create ``fan'' links to other users so they can easily follow each others' contributions.

\texttt{Digg} lets its users share news stories they discover from the Internet in the form of an URL. On this website each new submission immediately appears on a repository web page called ``upcoming stories'', where other members can find it and, if they like it, add a ``digg'' to it. A so-called ``digg number'' is shown next to each story's headline, which simply counts how many users have dugg the story in the past. If an upcoming story receives a sufficient number of diggs within a certain time period, it will be lifted from the pool of upcoming stories and be displayed on the front page, in which case it is said to be ``promoted'' or become ``popular''.\footnote{The actual proprietary algorithm used to determine whether a story qualifies to appear on the front page is more complex and will not be discussed in this paper.} A popular story normally keeps accumulating diggs for a few more days, before it eventually saturates (i.e.~stops receiving diggs) \cite{wu-huberman-07}. The final digg number of a story thus measures the total amount of attention it received in its lifetime.

Our first data set contains 2,676,160 stories submitted by 372,701 users between \texttt{Digg}'s inception in December 2004 and January 3, 2008. A total number of 59,853,763 diggs made on these stories, including their timestamps and submitter IDs, were fully recorded.

We also collected data from \texttt{Youtube.com}, a popular website where users can upload, view and share video clips. Similar to \texttt{Digg}, a ``view count'' number is displayed next to each video's title, measuring how many times it has been watched. However, in contrast with digg numbers which always saturate, this view count may increase {\it ad infinitum} \cite{gabor}.\footnote{There are at least two reasons why the digg number of news stories saturates but the view count of videos does not. First, videos are not as time-sensitive as news stories, so their novelty decays more slowly. Second, people search for videos on \texttt{Youtube} but browse through news stories on \texttt{Digg}, so it is easier to discover vintage videos than vintage stories.} Our second data set consists of 9,896,784 video uploads submitted by 579,470 \texttt{Youtube} contributors between the website's inception in February 2005 and May 1, 2008. For each video we recorded its submission date, uploader ID and ``final'' view count as of May 1, 2008.

As a counterpart of \texttt{Digg}'s ``popular stories'', \texttt{Youtube} also advertises a list of ``promoted/featured videos'' on its front page, which are chosen by the website staffs based on their personal tastes instead of an objective algorithm. Unfortunately in our data set we do not have the information which videos have been promoted. Therefore in order to calculate the ``popular ratio'' as we did for \texttt{Digg}, we had to resort to our own definition of ``popular videos''. We defined a video to be ``popular'' if and only if it received more than 10,000 views before May 1, 2007 when the data was collected, where 10,000 is the mean and 90\% quantile of all view counts in our data set. We also defined a video to be ``final'' if it is the last upload of its contributor and was uploaded three months before the data collection time (i.e.~uploaded before February 1, 2008).

\section{The active core and decreasing stopping probability}

\begin{figure}
\centering
\begin{minipage}{2.5in}
\centering
\includegraphics[width=2.5in]{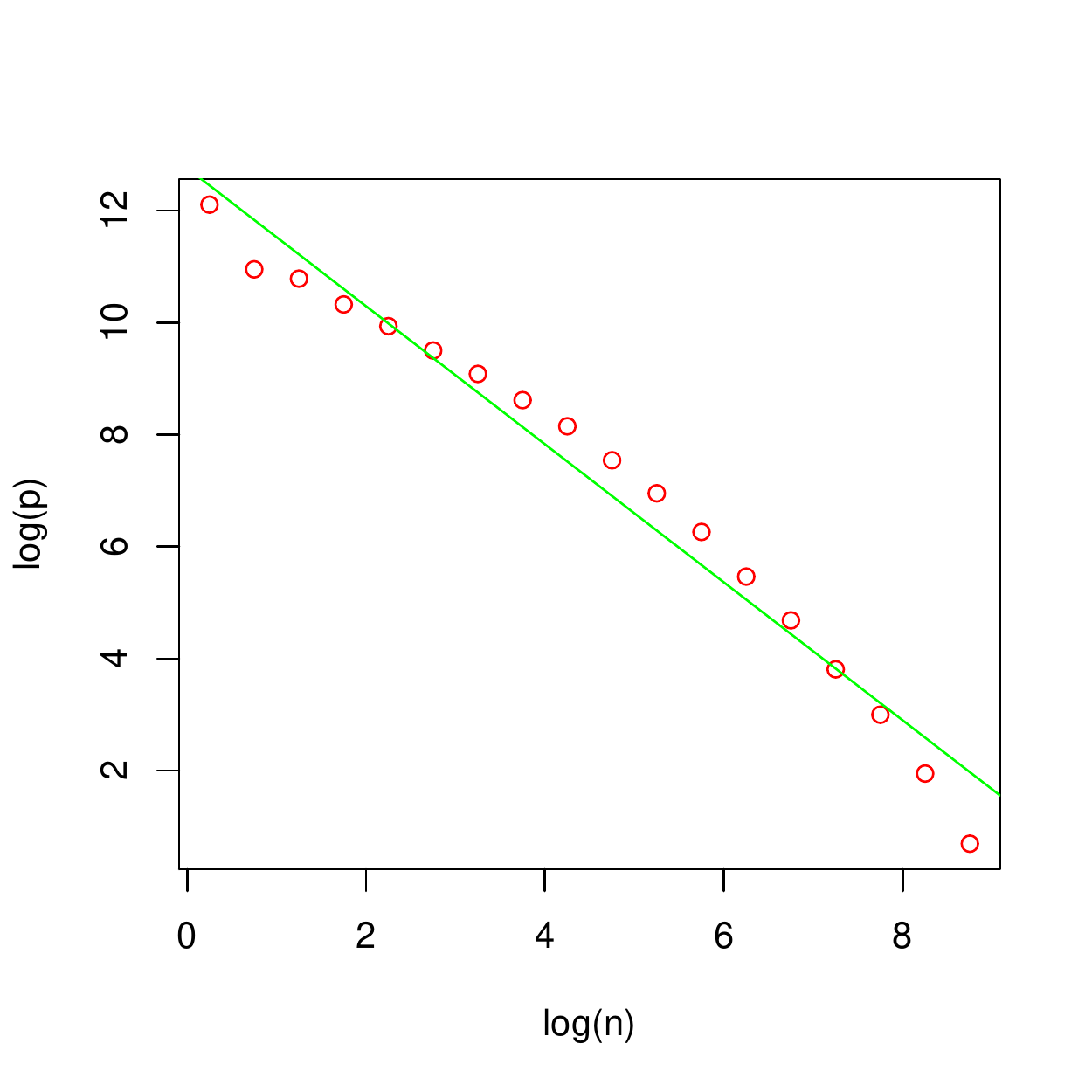}\\{\small (a)}
\end{minipage}
\begin{minipage}{2.5in}
\centering
\includegraphics[width=2.5in]{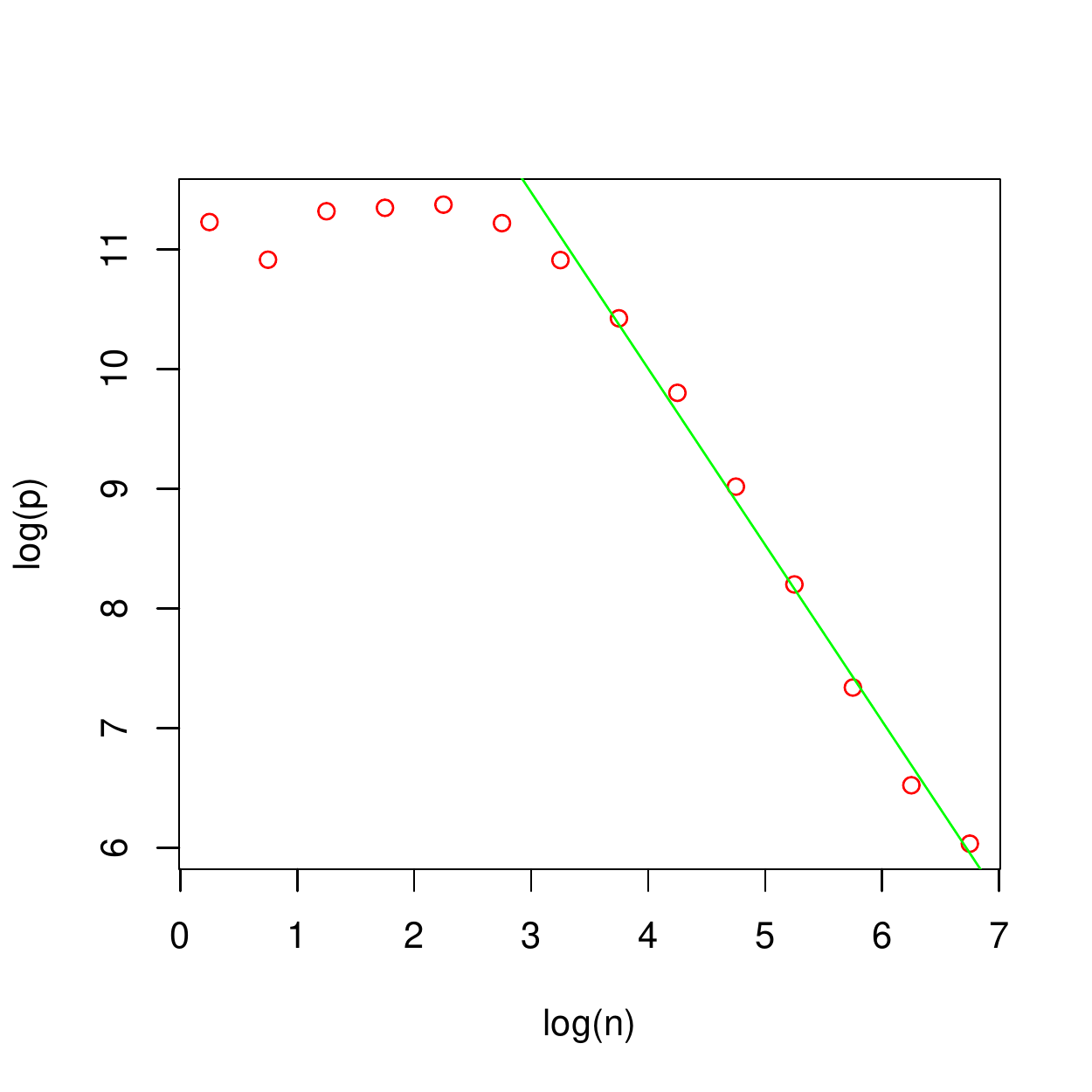}\\{\small (b)}
\end{minipage}
\caption{Distribution of the the number of submissions ($n$) for (a) \texttt{Digg} and (b) \texttt{Youtube}. The number of submissions follows a power-law distribution on \texttt{Digg}, and has a long tail on \texttt{Youtube}. It is evident that while the majority of users made relatively few contributions, there is also a considerable proportion of users who contributed substantially. For example, in our data sets 3,381 \texttt{Digg} users submitted more than 100 news stories, and 12,171 \texttt{Youtube} users uploaded more than 100 videos.}
\label{fig:n_submissions}
\end{figure}

We previously mentioned that a disproportionate number of contributions to online peer production efforts are made by a small number of very active users. This is true of \texttt{Digg} and \texttt{Youtube} as well.
As demonstrated in Fig.~\ref{fig:n_submissions}, the distribution of the number of contributions per user exhibits a heavy power-law tail. In this figure we include only those users who seem to have become inactive, by not contributing for at least $T$ months before the date of data capture, in order to accurately represent the ``final'' number of contributions for users. Of course, it is impossible to ascertain whether a given user has truly stopped or is merely taking an extended break, but in practice there may not be a large difference between these two states in terms of the user's morale and motivation. We selected $T=3$ in creating these plots; other choices produce similar results.

The power law in number of contributions per user is of interest in our context not only because it indicates the presence of an active core of participants, but also because of what it implies about users' motivation. Given a large number of users, the proportion that stop after making their $n$'th contribution is
\[
P(\mbox{stop after}~n) = \frac{N(n)}{\int_n^{n_\mathrm{max}} N(m) dm},
\]
where $N(n)$ is the number of users making $n$ contributions and then stopping and $n_\mathrm{max}$ is some empirical maximum number of contributions. Empirically, at the high end, $N(n)$ follows a power law form, $N(n) \sim n^{-k}$ with $k>1$, so therefore
\[
P(\mbox{stop after}~n) \sim 1/n + (n/n_\mathrm{max})^{-k+1}.
\]
We thus see that the probability a user stops contributing decreases the more he contributes \cite{wilkinson-08}.

\section{Attention and the role of popularity}
As an alternative to user-cited motivating factors listed above, we propose that contributors to \texttt{Digg} and \texttt{Youtube} are motivated by attention. We first provide empirical evidence of this, and then  describe how attention a user receives tends to keep increasing via a follower reinforcement mechanism, thereby explaining the power law in contributions per user.

\subsection{Lack of attention is correlated to stopping}
\begin{figure*}
\centering
\includegraphics[width=5in]{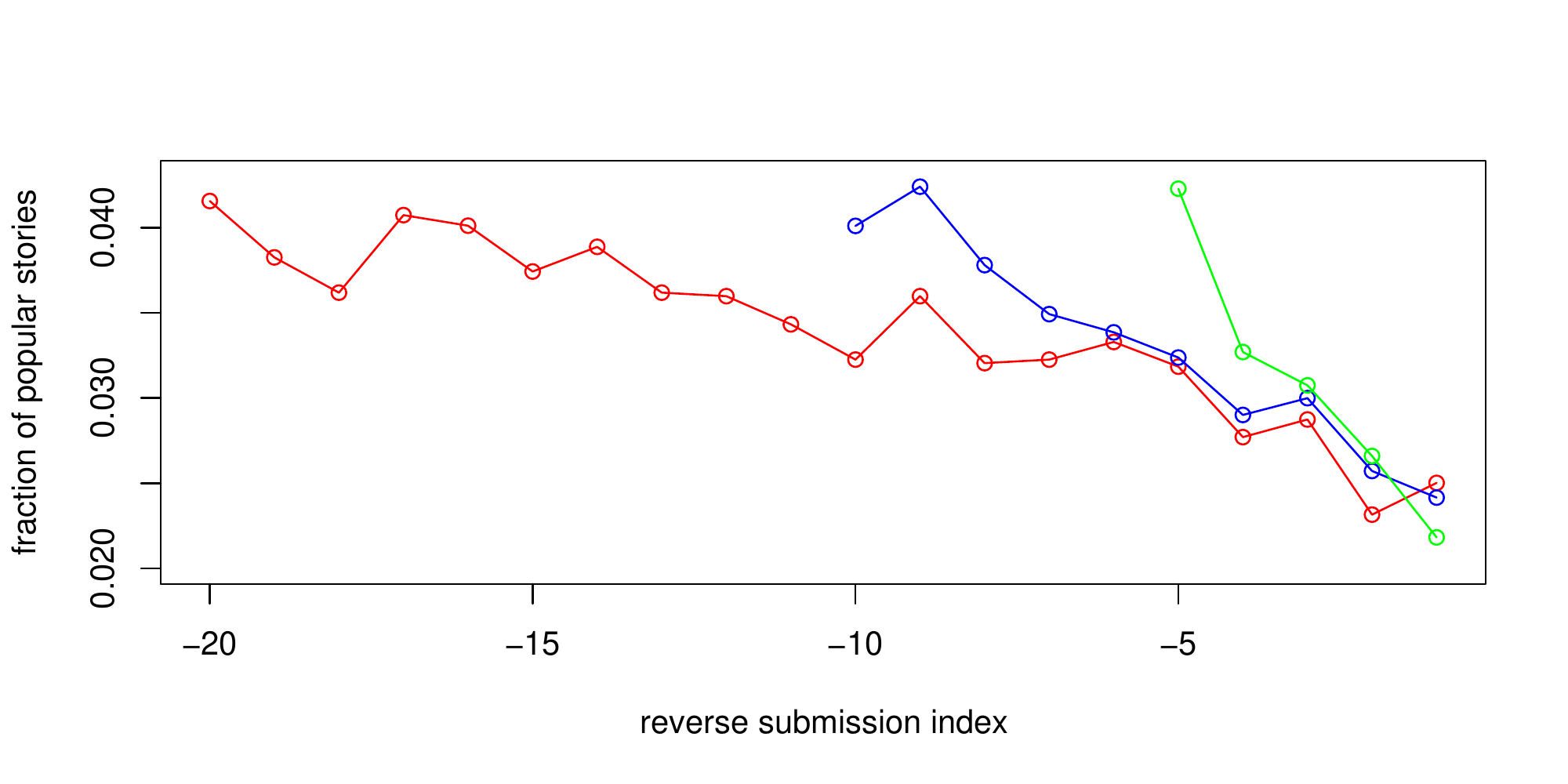}
\caption{Popular ratio versus reverse submissions index for \texttt{Digg} contributors. The horizontal axis marks the reverse index of a user's submission, so that $-1$ is her last submission, $-2$ is her second last submission, and so on. The green, blue and red curves are calculated for those users who submitted no less than 5, 10 and 20 stories in their lifetime, respectively.}
\label{fig:digg_popularity}
\end{figure*}

If \texttt{Digg} contributors are motivated by attention, we would expect that story submitters whose stories do not achieve popularity be more likely to stop than others. Again, we use a threshold of 3 months of inactivity to demarcate stopping; even if a user returns after a long hiatus, he most likely was quite unmotivated during his time away. We thus chose the 30,157 users who submitted no less than 5 stories before they stopped (i.e.~submitted at least 5 stories before October 3, 2007 and had no activity since then till January 3, 2008, when our data was collected), and calculated the fraction of popular stories among all the 30,157 $n$'th last stories, for $n=1,2,3,4,5$. The result is plotted in Fig.~\ref{fig:digg_popularity} as the green curve. As one can see more than 4\% of all the 5th last stories eventually became popular, but only 2\% of the last stories had the same luck. This is consistent with our conjecture that a contributor is more likely to stop when she fails to receive enough attention. To show that our result is robust, we repeated the calculation for two more groups of users who submitted no less than 10 and 20 stories in their lifetime. The fraction of popular stories again exhibits declining trends for both groups, as can be seen from the blue and red curves in Fig.~\ref{fig:digg_popularity}.

\begin{figure*}
\centering
\includegraphics[width=5in]{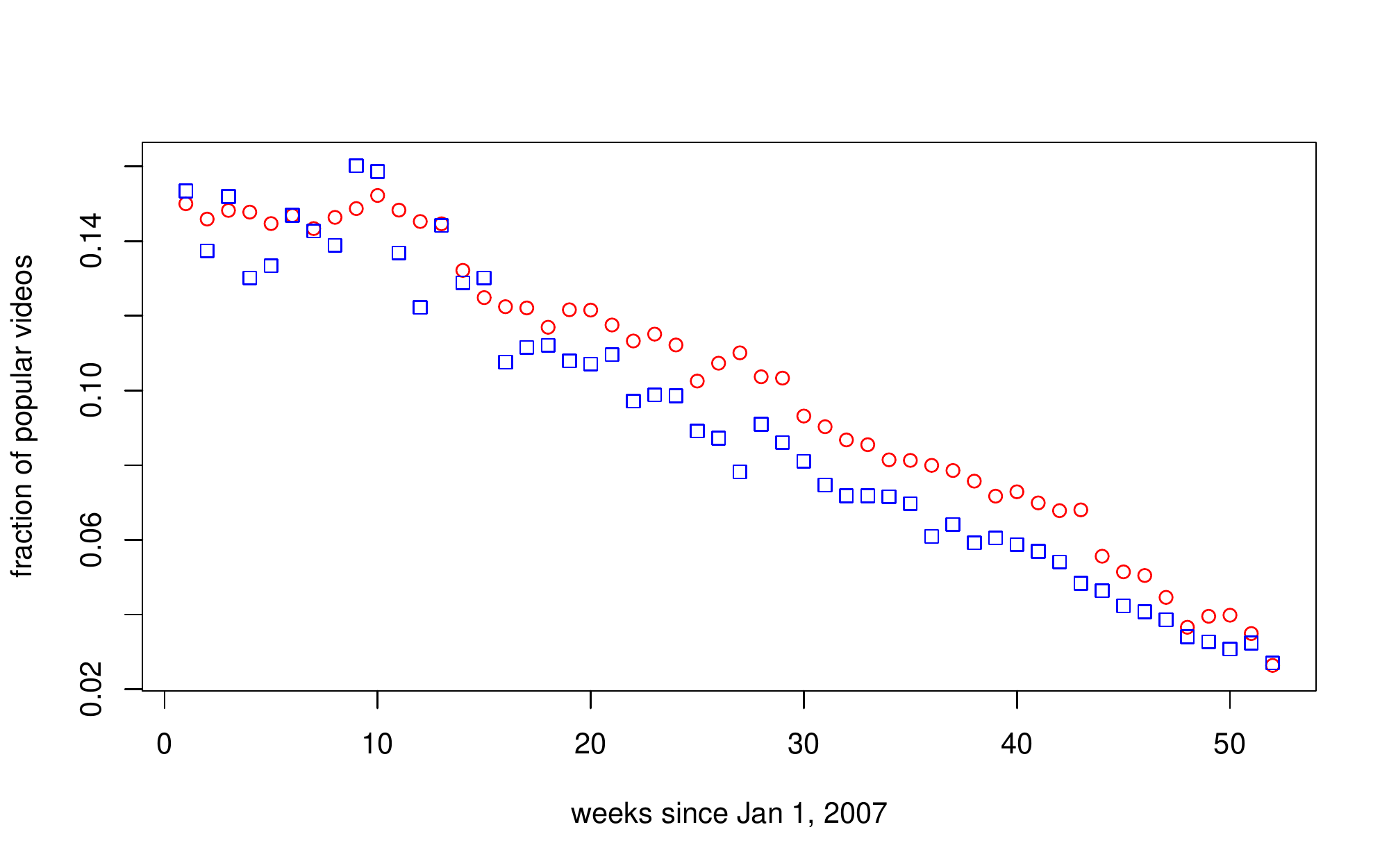}
\caption{The red circles in this figure represent the popular ratios of \texttt{Youtube} videos submitted in each week of 2007. The blue squares represent the popular ratios of all final submissions made in the same weeks. As can be seen most squares lie below the circles, indicating that the final submissions received less attention than the earlier ones.}
\label{fig:youtube_popularity}
\end{figure*}

We then calculated the popular ratio of all \texttt{Youtube} videos submitted in each week of the year 2007 (denoted by $r(t)$), as well as the popular ratio of all final videos submitted in the same weeks (denoted by $r_f(t)$), where popularity was defined in Section \ref{sec:data}.\footnote{\label{foot:youtube_growth}Because the view count of \texttt{Youtube} videos increases infinitely, late submissions generally receive less view counts than earlier ones. It no longer makes sense to plot the fraction of popular videos as a function of the video's reverse submission index, as we did in the study of \texttt{Digg} (Fig.~\ref{fig:digg_popularity}). This is because one would not able to tell whether the final videos received less view counts because they are less popular or simply because they have been on the website for a shorter time. Hence it is only meaningful to compare the view counts of videos submitted not far apart in time.} Finally, we plotted the two sets of ratios in Fig.~\ref{fig:youtube_popularity}. As one can see the final videos on average are less popular than the other videos submitted in the same weeks, which is consistent with our hypothesis that lack of popularity triggers cessation. A paired $t$-test of the alternative hypothesis ``$r_f(t)<r(t)$'' yields a $p$-value less than 0.001, again supporting our hypothesis.

\subsection{Attention tends to increase with the number of submissions}
The evidence presented above suggests a strong correlation between popularity (attention received) and productivity (tendency to contribute). It is thus natural to ask whether one could develop a model to predict a user's productivity from her popularity. Suppose that the amount of attention each user receives from one  submission is a random variable $X$. Consider for benchmark purpose the following simplest stopping rule. If a contributor's latest submission surpasses some ``successful'' threshold $\theta$ (which may also be random and may be a different random variable for each person) she will continue to contribute, otherwise she will stop.\footnote{A more realistic model would assume that a user's stopping probability depends not only on the popularity of her last submission but also on her earlier ones. This will not change the following discussion in a fundamental way.} Under these assumptions it is straightforward to show that the number of user submissions $n$ will follow a geometric distribution: $p_n = p(1-p)^{n-1}$ for $n\ge 1$, where $p=P(X>\theta)$. This, however, is at odds with the power-law distribution of user productivity which exhibits a long tail.

The reason for this discrepancy lies in our assumption that the amount of attention received by each submission is always drawn from the same distribution, so that each submission has an equal chance to succeed. If we assume instead that the attention received by each contributor is positively reinforced as time goes on, then those who have been contributing for a longer time will be more likely to obtain a ``successful'' status with their new submissions, thus less likely to stop. Such a modification would lead to a tail distribution of user submissions that is longer than the geometric tail and closer to the power-law tail that has been observed on \texttt{Digg}, \texttt{Youtube} and many other websites.

\begin{figure}
\centering
\begin{minipage}{2.5in}
\centering
\includegraphics[width=2.5in]{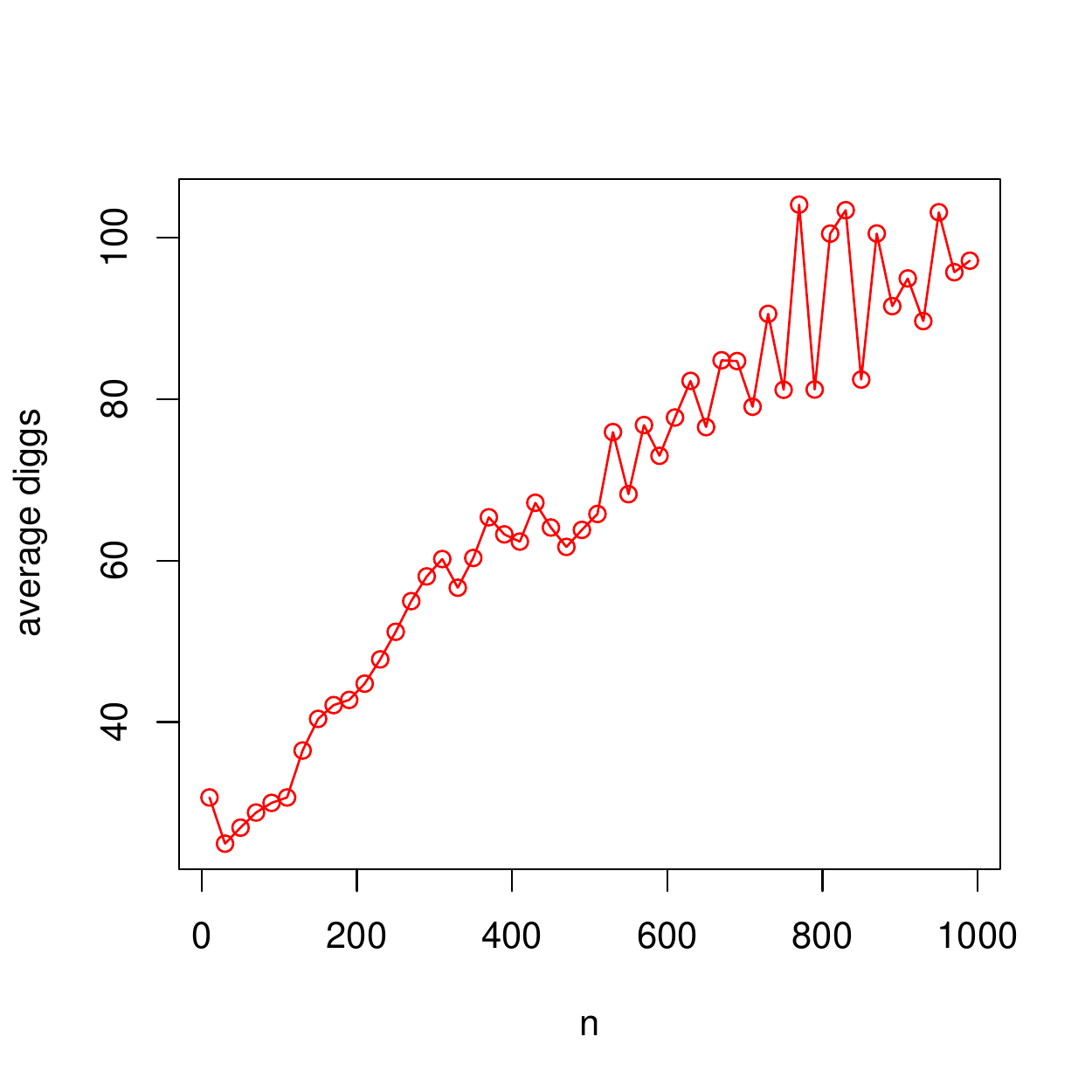}\\{\small (a)}
\end{minipage}
\begin{minipage}{2.5in}
\centering
\includegraphics[width=2.5in]{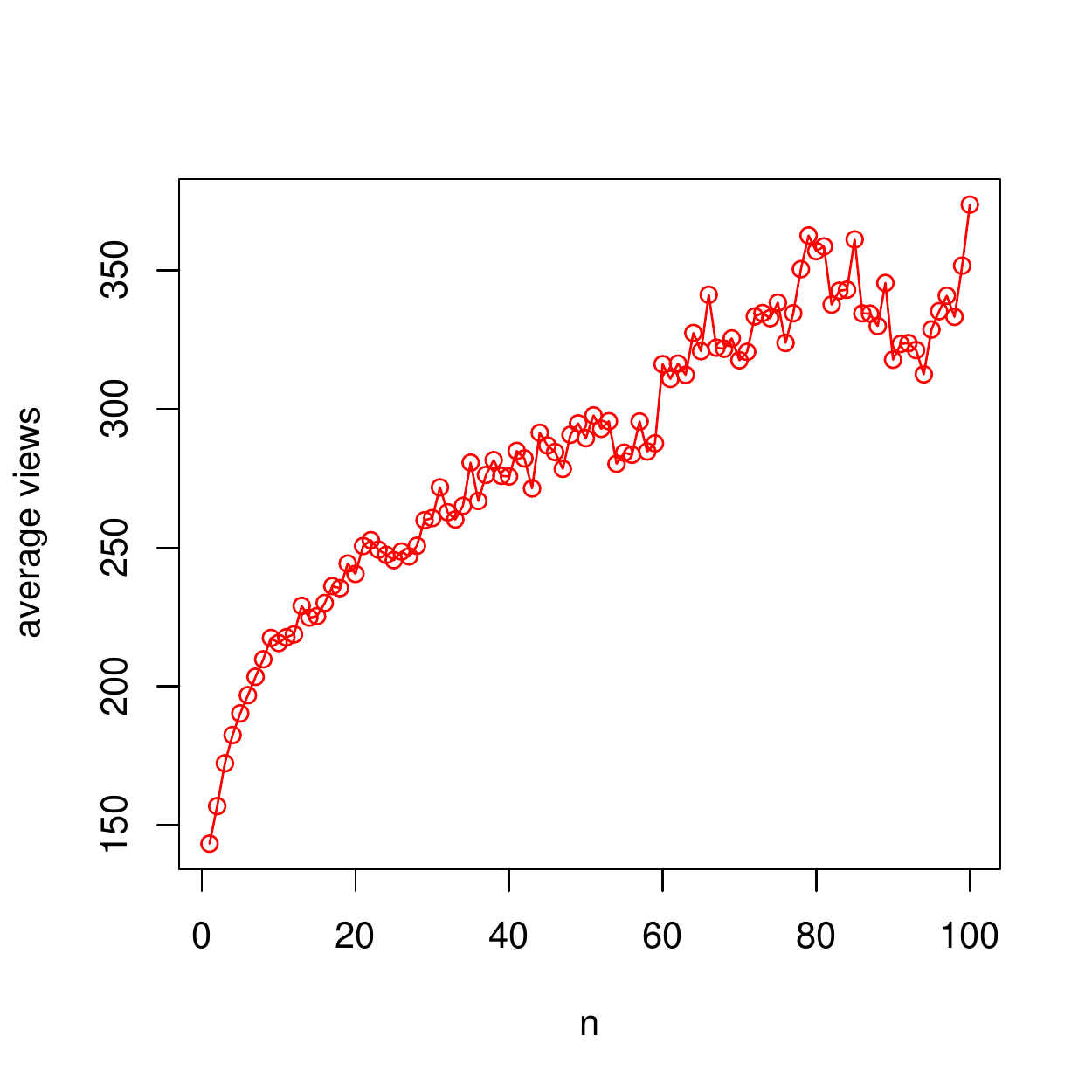}\\{\small (b)}
\end{minipage}
\caption{The amount of attention received by each submission as a function of the submission index. The $n$'th point in each figure plots the average amount of attention (digg number or view count) received by all submissions whose index lies between $20(n-1)$ and $20n$, right inclusive. In plotting (a) we made use of the full \texttt{Digg} data set with the removal of four contributors who had abnormally many fans. They are \texttt{ObamaforAmerica} (Barack Obama, 9896 fans and 49 submissions), \texttt{teamhillary} (Hillary Clinton, 1305 fans and 14 submissions), \texttt{Gravel08} (Mike Gravel, 2609 fans and 3 submissions), and \texttt{jayadelson} (Jay Adelson, CEO of \texttt{Digg}, 1956 fans and 40 submissions). In plotting (b) we filtered out all videos submitted in January 2007 from the \texttt{Youtube} data set (c.f.~footnote \ref{foot:youtube_growth}).}
\label{fig:popularity_vs_index}
\end{figure}

To verify this reinforcement theory we plotted the amount of attention received by each submission as a function of the submission index, for both \texttt{Digg} and \texttt{Youtube}. The results are shown in Fig.~\ref{fig:popularity_vs_index}. It can be seen that the received attention $X$ grows more or less proportionally with the submission index $n$, which confirms the existence of attention reinforcement. Mathematically, we can write $X=anY$ where $a$ is a positive constant and $Y$ is a multiplicative noise with mean one.\footnote{We assume multiplicative noise instead of additive noise because there is empirical evidence that attention growth is a multiplicative process \cite{wu-huberman-07}.} Assuming the same stopping rule as before, a new submission with index $n$ will fail to be successful with probability $P(anY<\theta) = P(aY/\theta < 1/n)$. Let $F$ be the CDF of the random variable $aY/\theta$. We see that a contributor who made $n$ past submissions will stop at $n$ with probability
\begin{equation}
F\left(\frac 1n\right) = F(0) + \frac{F'(0)} n + O\left(\frac 1{n2}\right) = \frac{F'(0)} n + O\left(\frac 1{n2}\right),
\label{eq:stop}
\end{equation}
where $F'(0)$ is a positive constant. Suppose the number of contributions in the population follows a distribution $p_n$, or equivalently, a fraction $p_n$ of contributors stop at $n$ contributions. Using this notation and ignoring the higher-order terms for large $n$, Eq.~(\ref{eq:stop}) can be written as
\begin{equation}
\frac{p_n}{p_n+p_{n+1}+\cdots} = \frac{F'(0)}n.
\end{equation}
Defining $G(n)=p_n+p_{n+1}+\cdots$ and approximating $G(n)$ with a continuous function, we have
\begin{equation}
-\frac{G'(n)}{G(n)} = \frac{F'(0)}n.
\end{equation}
The solution to this equation is $G(n)\sim n^{-F'(0)}$, which implies that $p_n=-G'(n) \sim n^{-F'(0)-1}$. We have thus shown that the number of contributions has a power-law tail, in accordance with the empirical observations.

\subsection{Attention is reinforced by one's followers}

While our reinforcement theory offers a tentative explanation for the long tail distribution of user contribution, it does not tell us why such reinforcement exists. It is possible that the contributors learn by experience what is favored by the general population and adapt their contributions to the popular demand. This would explain why later submissions have higher popularity, but not the linear dependency (Fig.~\ref{fig:popularity_vs_index}). To answer this question one needs to look into the constituents of a contributor's potential audience.

As a matter of fact, previous empirical studies have shown that the majority of diggs received by a \texttt{Digg} story in the upcoming phase (i.e.~before it is promoted to the front page) come from the contributor's ``fans''. This is due to the fact that \texttt{Digg} implemented a \emph{social filtering system} that allows its users to designate other users as ``friends'' and easily track friends' activities \cite{lerman-07}. When a contributor submits a story to \texttt{Digg}, her fans (those who designated the contributor as friend, also called ``reversed friends'') can discover the story through a friend interface and digg it if they like it. On \texttt{Youtube} there is a similar ``subscription'' mechanism that allows its users to subscribe to other users' videos. When a contributor releases a new video it immediately become visible to her fans (those who subscribed to her videos) through a subscription interface.

Because a considerable portion of attention a contributor receives can be attributed to her fans, the contributor's publicity (measured by the number of fans) could act as the important missing link between popularity and productivity. A contributor with many past contributions (high productivity) naturally has many fans (high publicity). Her fans naturally pay a lot of attention to her next contribution (high popularity). This in turn incentivizes the contributor to make more contributions (higher productivity), thereby closing the reinforcement loop.

\begin{figure}
\centering
\begin{minipage}{2.5in}
\centering
\includegraphics[width=2.5in]{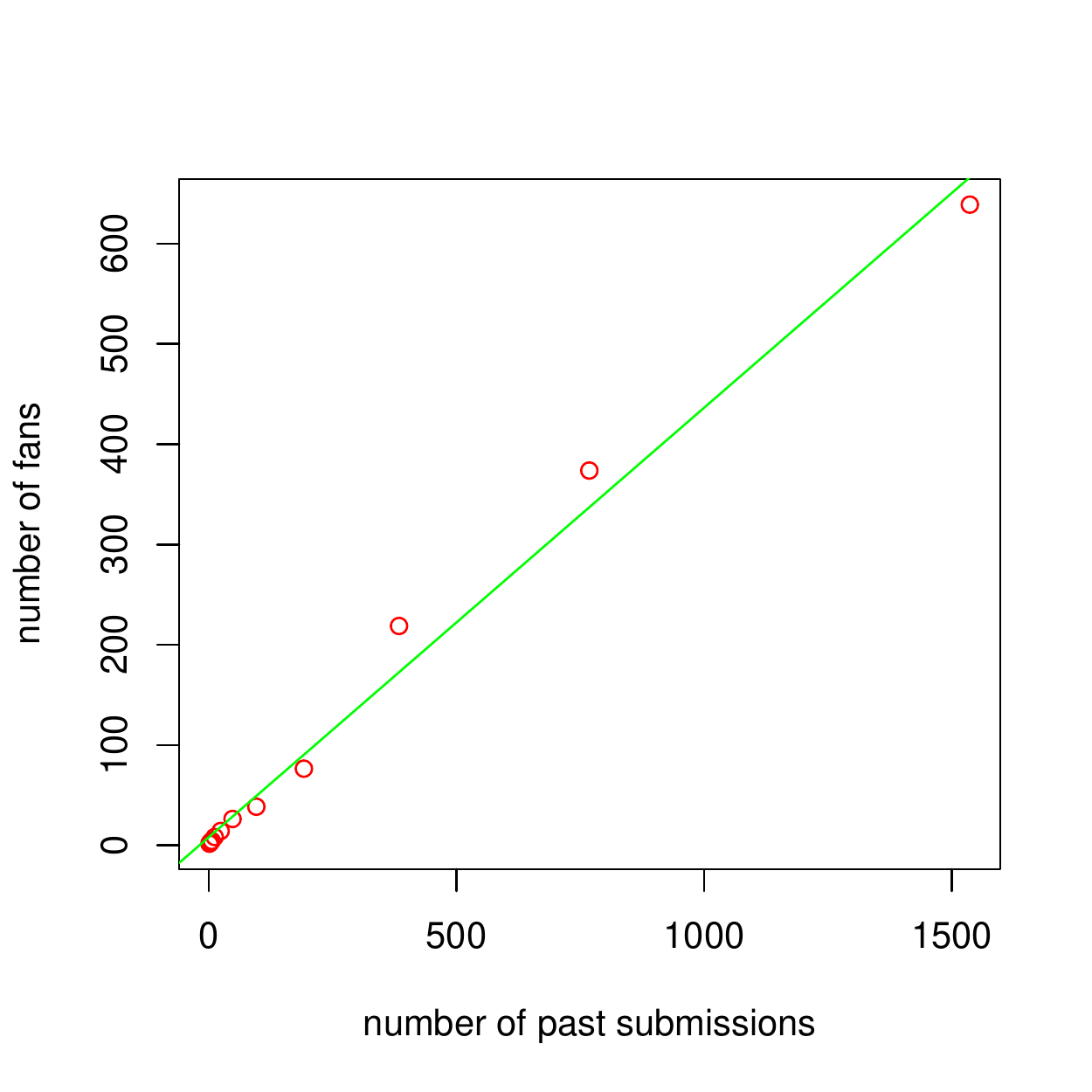}\\{\small (a)}
\end{minipage}
\begin{minipage}{2.5in}
\centering
\includegraphics[width=2.5in]{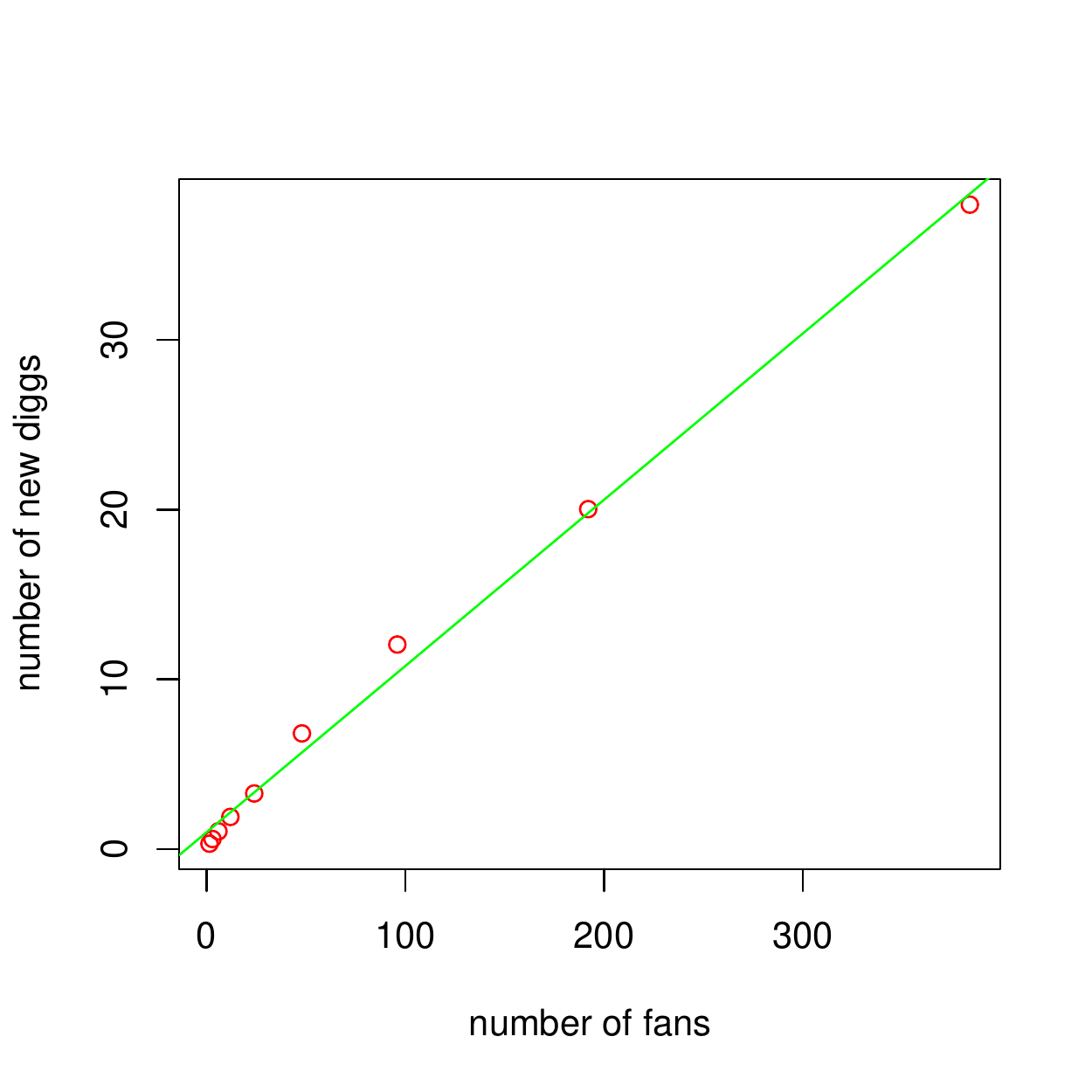}\\{\small (b)}
\end{minipage}
\caption{We recorded the digg number of all 145,081 stories submitted by 30,341 \texttt{Digg} contributors between January 3, 2008 and January 17, 2008, as well as the contributors' fan number on January 3, 2008. (a) Publicity (number of fans) as a function of productivity (number of past contributions). The $n$'th data point plots the average fan number of those contributors who submitted $2^{n-1}\le m <2^n$ stories before January 3, 2008. (b) Popularity (digg number) as a function of publicity (number of fans). The $n$'th data point plots the average digg number of all stories whose contributor had $2^{n-1}\le m <2^n$ fans on January 3, 2008.}
\label{fig:digg_fans}
\end{figure}

To test the role of publicity we recorded the fan number of each contributor in our \texttt{Digg} data set on January 3, 2008. We then tracked the digg number of all stories submitted in the next two weeks (between January 3, 2008 and January 17, 2008). A total number of 145,081 such stories were submitted by 30,341 contributors. Fig.~\ref{fig:digg_fans}(a) plots the contributors' publicity (number of fans on January 3) as a function of their productivity (number of contributions before January 3), and Fig.~\ref{fig:digg_fans}(b) plots their popularity (final digg number of each new contribution submitted between January 3 and January 17) as a function of their publicity. It can be seen from the two figures that a contributor's popularity is roughly proportional to her publicity, which in turn is roughly proportional to her productivity. Summarizing the two results, this explains why popularity grows linearly with productivity, which is what we found from Fig.~\ref{fig:popularity_vs_index}(a).

\begin{figure}
\centering
\begin{minipage}{2.5in}
\centering
\includegraphics[width=2.5in]{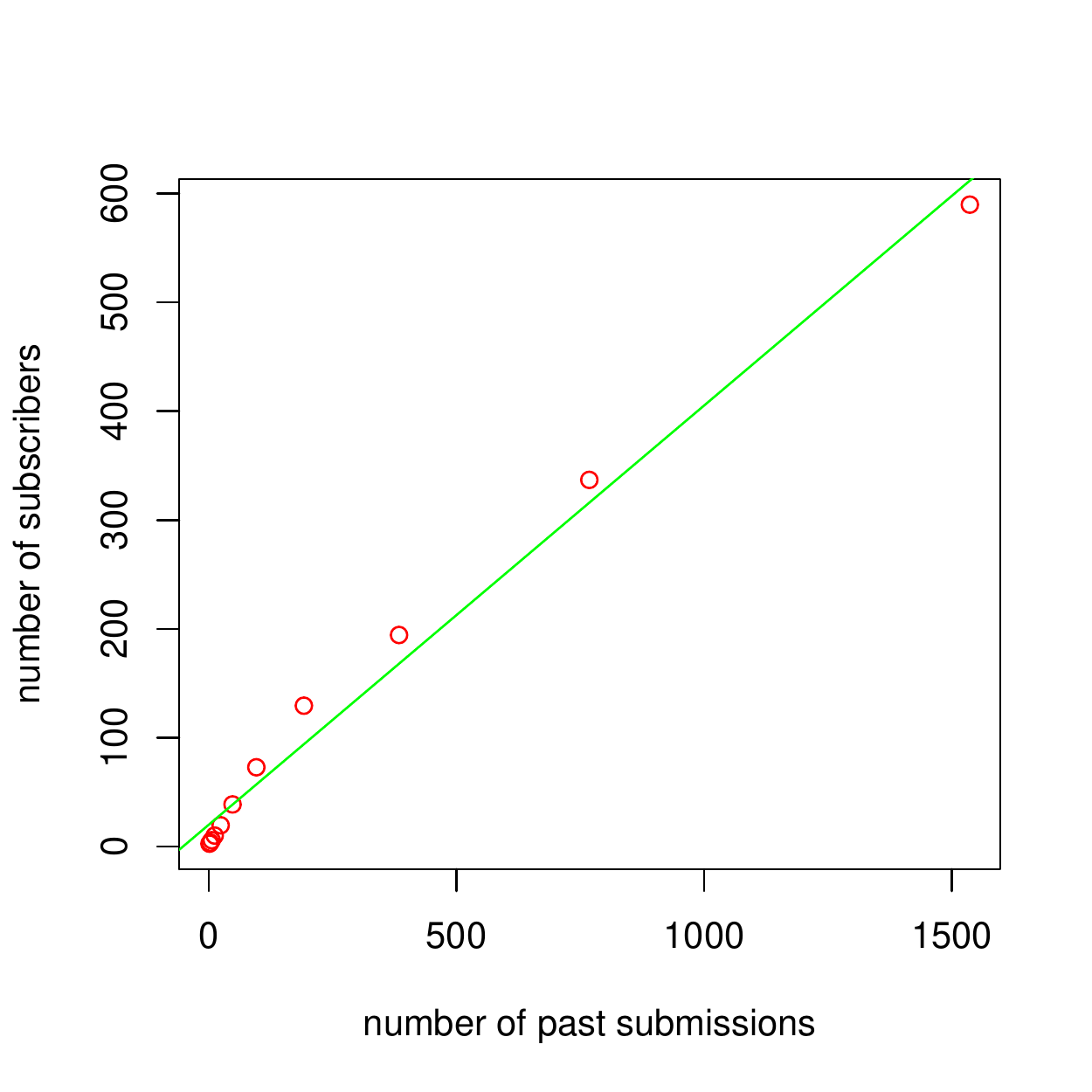}\\{\small (a)}
\end{minipage}
\begin{minipage}{2.5in}
\centering
\includegraphics[width=2.5in]{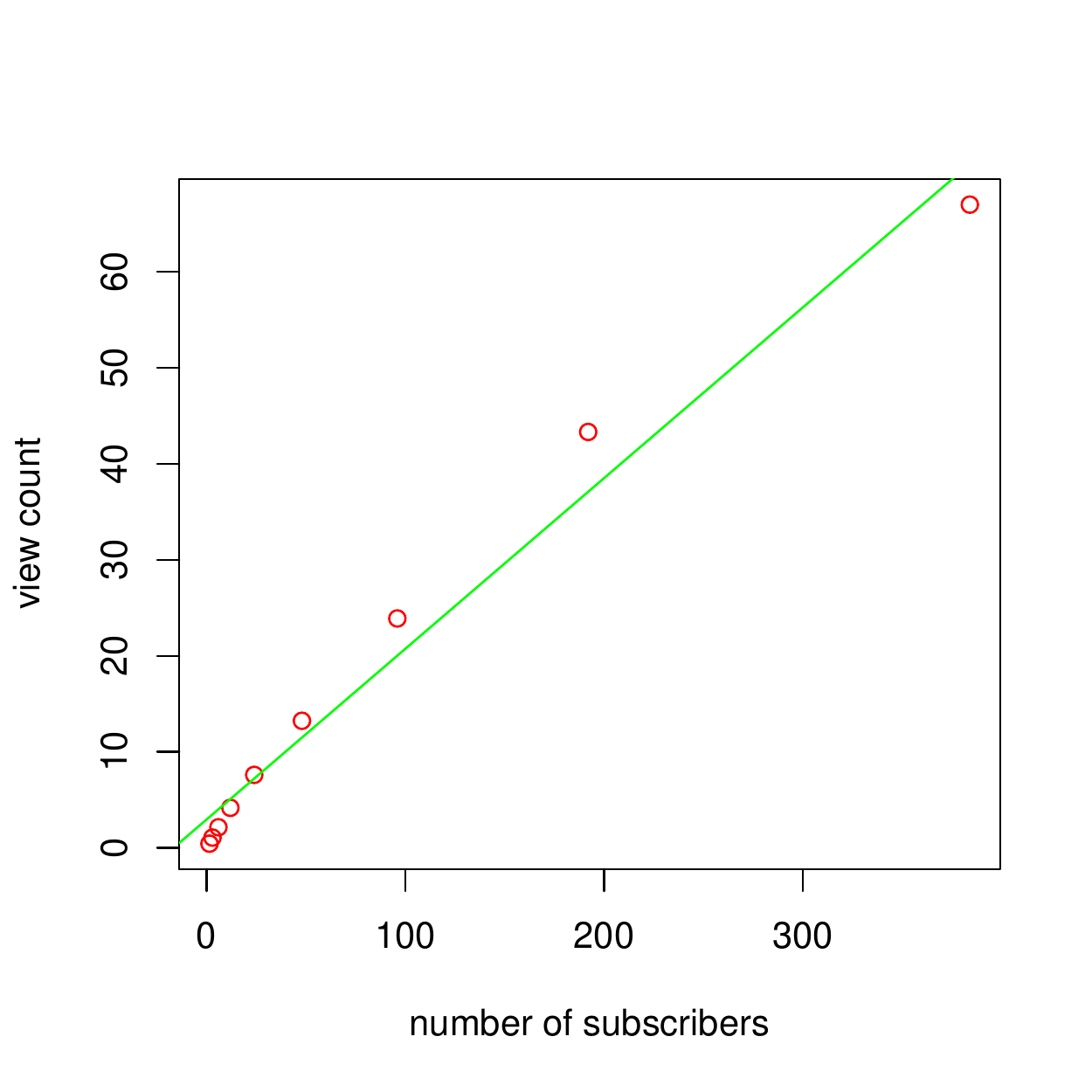}\\{\small (b)}
\end{minipage}
\caption{We recorded the view count of 260,002 videos submitted by 86,620 \texttt{Youtube} contributors between February 7, 2008 and February 21, 2008, as well as the contributors' subscription number on February 7, 2008. (a) Publicity (number of subscribers) as a function of productivity (number of past contributions). The $n$'th data point plots the average subscription number of those contributors who uploaded $2^{n-1}\le m <2^n$ videos before February 7, 2008. (b) Popularity (digg number) as a function of publicity (number of subscribers). The $n$'th data point plots the average view count of all videos whose contributor had $2^{n-1}\le m <2^n$ subscribers on February 7, 2008.}
\label{fig:youtube_fans}
\end{figure}

We next performed a similar measurement on \texttt{Youtube}. On February 7, 2008 we recorded the subscription number of each contributor in our \texttt{Youtube} data set. A total number of 86,620 contributors submitted 260,002 videos in the following two weeks (between February 7, 2008 and February 21, 2008). We plotted the average subscription number of these contributors as a function of their past productivity (Fig.~\ref{fig:youtube_fans}(a)), and also plotted the average view count each contributor received in the two studied weeks as a function of their subscription number (Fig.~\ref{fig:youtube_fans}(b)). Again we observed proportionality in both figures.

\section{Conclusion}
In this paper we have shown that attention is a likely motivating factor for the active core of participants to two popular peer production websites. Followers affect contributions to peer production websites, and motivate prolific contributors to remain active for a long time. We demonstrated a strong correlation between lack of attention and users' decision to stop contributing in both youtube.com and social news site digg.com. We then showed that the more users contribute, the greater their following (as measured by their number of digg ``fans'' or youtube ``subscribers''), and the more attention they tend to receive from these followers. Finally, we presented a mechanism incorporating these empirical findings which explains the observed power law in the distributions of contributions per user. Taken together, our findings suggest that the feedback loop of attention is an important ingredient in a successful of a peer production effort because of its role in motivating devoted contributors to persist.


\begin{thebibliography}{99}
\bibitem{wilkinson-08} D. M. Wilkinson. Strong regularities in online peer production. \emph{ACM Conference on Electronic Commerce}, 2008.

\bibitem{shah} S. K. Shah. Motivation, Governance, and the Viability of Hybrid Forms in Open Source Software Development. \emph{Management Science} vol. 52, no. 7, pp. 1000--1014, 2006

\bibitem{lakhani} K. Lakhani and R. G. Wolf. Why hackers do what they do: Understanding motivation and effort in free/open source software projects. In \textit{Perspectives on Free and Open Source Software}, Joe Feller, Brian Fitzgerald, Scott Hissam and Karim Lakhani, eds. Cambridge, Mass: MIT Press, 2005.

\bibitem{nov} O. Nov. What motivates Wikipedians? \emph{Communications of the ACM},
vol. 50, no. 11, pp. 60--64, 2007.

\bibitem{hsu} C.-L. Hsu and J. Lin. Acceptance of blog usage: The roles of technology acceptance, social influence and knowledge sharing motivation. \emph{Information and Management}, vol. 45, no. 1 pp. 65--74, 2008.

\bibitem{oreg} S. Oreg and O. Nov. Exploring Motivations for Contributing to Open Source Initiatives: The Roles of Contribution Context and Personal Values. \emph{Computers in Human Behavior}, vol. 24, no. 5, pp. 2055--2073, 2008.

\bibitem{roberts} J. Roberts, I.-H. Hann, and S. Slaughter. Understanding the Motivations, Participation and Performance of Open Source Software Developers: A Longitudinal Study of the Apache Projects. \emph{Management Science}, vol. 52, no. 7, pp. 984--999, 2006.

\bibitem{lampel} J. Lampel and A. Bhalla. The role of status seeking in online communities: Giving the gift of experience. In \emph{Journal of Computer-Mediated Communication}, vol. 12, no. 2, article 5, 2007.

\bibitem{okoli} C. Okoli and W. Oh. Investigating recognition-based performance in an open content community: A social capital perspective. \emph{Information and Management}, vol. 44, no. 3, pp. 240--252, 2007.


\bibitem{franck} G. Franck. Science communication, a vanity fair. \emph{Science}, vol. 286, pp. 53--55, 1999.

\bibitem{huberman-04} B. A. Huberman, C. Loch and A. Onculer. Status as a valued resource. \emph{Social Psychology Quarterly}, vol. 67, no. 1, pp. 103--114, 2004.

\bibitem{romero} B. A. Huberman, D. M. Romero and F. Wu. Crowdsourcing, attention, and productivity. Under review.

\bibitem{miura} A. Miura and K. Yamashita. Psychological and social influences on blog writing: An online survey of blog authors in Japan. \emph{Journal of Computer-Mediated Communication}, vol. 12, no. 4, article 15, 2007.

\bibitem{wu-huberman-07} F. Wu and B A. Huberman. Novelty and collective attention. \emph{Proc.~Natl.~Acad.~Sci.}, vol.~104, no.~45, pp.~17599--17601, 2007.

\bibitem{gabor} G. Szabo and B. A Huberman. Predicting the popularity of online content. Under Review.
by: 
\bibitem{lerman-07} K. Lerman. Social networks and social information filtering on \texttt{Digg}. \emph{Proceedings of International Conference on Weblogs and Social Media (ICWSM)}, 2007.

\end{thebibliography}
\end{document}